\documentclass[12pt,a4]{article}

\usepackage[square,numbers, sort&compress]{natbib}

 \usepackage{hyperref}
\hypersetup{colorlinks=true, citecolor=blue, urlcolor=blue, linkcolor=blue}

\usepackage{authblk}

\usepackage{color}
\usepackage{graphicx}

\usepackage{cancel}
\usepackage[normalem]{ulem}

\usepackage{amsmath,amssymb}

\usepackage[utf8]{inputenc}

\usepackage{bm}

\usepackage{mathptmx}
\usepackage{etoolbox}
\usepackage[normalem]{ulem}
\usepackage{xcolor}

\begin{document}

\title{Learning by mistakes in memristor networks} 


\author[1]{Juan Pablo Carbajal}
\affil[1]{Institute for Energy Technology,  University
of Applied Sciences of Eastern Switzerland, Oberseestrasse 10, 8640 Rapperswil, Switzerland}
\author[2]{Daniel A. Martin}
\affil[2]{Center for Complex Systems and Brain Sciences (CEMSC${^3}$) \& Instituto de Ciencias Físicas (ICIFI-Conicet), Escuela de Ciencia y Tecnología, Universidad Nacional de Gral. San Martín, Campus Miguelete, 25 de Mayo y Francia (1650), San Martín, Buenos Aires, Argentina}
\author[3]{Dante R. Chialvo}
\affil[3]{Center for Complex Systems and Brain Sciences (CEMSC${^3}$) \& Instituto de Ciencias Físicas (ICIFI-Conicet), Escuela de Ciencia y Tecnología, Universidad Nacional de Gral. San Martín, Campus Miguelete, 25 de Mayo y Francia (1650), San Martín, Buenos Aires, Argentina}

\date{\today}

\maketitle

\begin{abstract}
Recent results  
revived the interest in the implementation of analog devices able to perform brain-like operations.
Here we introduce a training algorithm for a memristor network which is inspired in previous work on biological learning.
Robust results are obtained from computer simulations of a network of voltage controlled memristive devices.
Its implementation in hardware is straightforward, being scalable and requiring very little peripheral computation overhead.
\end{abstract}

 {\section{Introduction}}
In the last decade we have witnessed an explosion in the interest on neuro-morphing, i.e.,   adaptive devices inspired in brain principles.
Many of the current efforts focus on the replication of the dynamics of a single neuron, using a diversity of technologies including magnetic, optics, atomic switches, etc~\cite{Review1}.
While the emulation of a single neuron seems achievable with present technology, we still lack learning algorithms to train large interconnected neuron-like elements, without resorting to \emph {peripheral computation overheads.}

In thinking about this issue, it is soon realized that {\it in vivo} biological learning exhibits important features which are not presently considered in neuromorphing implementations.
The most relevant one, in the context of these notes, is the fact that the only information a biological neuron has at its disposal to modify its synapses is either {\it global} or {\it local}.
In other words, in real brains, there is no \emph {peripheral computation overheads}, the strength of the synaptic weight between any two given neurons, is a function of the activity of its immediate neighbors and/or (through some so-called neuro-modulators) some global state of the brain (or partial region of), resulting from success or frustration in achieving some goal (or being happy, angry, excited, sleepy, etc).
These observations leaded to the proposal~\cite{bak,bak1,bak2,joe1,joe2,brigman} of a simple neural network model able to learn simple input/output associations.

The present article instantiates a translation of the work of Refs~\cite{bak,bak1} into the realm of memristive networks.
The main objective is to design a device working on the principles described there, able to be fully implemented in hardware, requiring no access to the inner structure of the network and minimal (i.e., one ammeter, one switch, and two batteries) external processing.

This article is organized as follows.
First 
we will review previous work~\cite{bak,bak1,bak2,joe1,joe2,brigman} describing a self-organized process by which biological learning may proceed.
Such work is the inspiration for the algorithm proposed here to train a network of memristors~\cite{chua,ReviewCarvelli} which is introduced after that. Subsequently
the  main results describing the simulation results obtained from a three layer feedforward network are presented.
The paper closes with a short list of expected hurdles to surpass and other possible similar implementations.
Numerical details are described on the Appendix, together with some miscellaneous observations.
\\
\\
\section{Algorithm, model and observables}
\label{Model}

 {\subsection{A toy model of biological learning}}

Two decades ago, Chialvo \& Bak~\cite{bak1} introduced an unconventional model of learning which emphasized self-organization.
 In that work, they re-examined the commonly held view that learning and memory necessarily require potentiation of synapses.
Instead, they suggested that, for a naive neuronal network, the process of learning involves making more mistakes than successful choices, thus the process of adapting the synapses would have more opportunities to punish the mistakes than to positively reinforce the successes.
Consequently, their learning strategy used two steps: the first involves extremal dynamics to determine the propagation of the activity through the nodes and the second using synaptic depression to decrease the weights involved in the undesired (i.e., mistaken) outputs.
The first step implies to select only the strongest synapses for propagating the activity.
The second step assumes that  active synaptic connections are temporarily tagged and subsequently depressed if the resulting output turns out to be unsuccessful.
Thus, all the synaptic adaptation leading to learning is driven only by the mistakes.

\begin{figure}  
\centerline{
\includegraphics [width = .75 \linewidth] {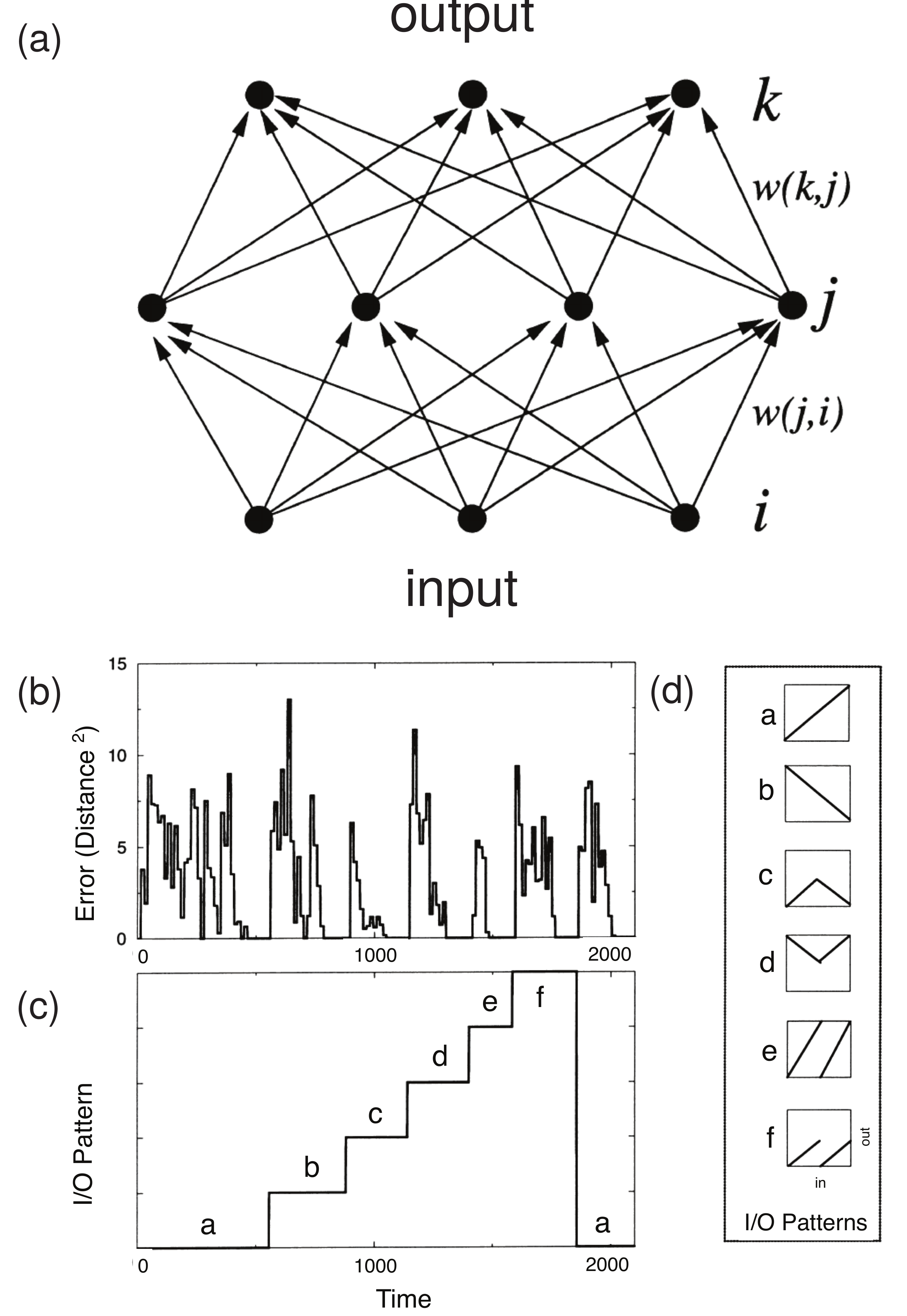} }
\caption{ \label{fig1} An example of a three-layer network with 3 input nodes, 4 intermediary nodes and 3 output nodes is shown in (a).
Each input node has a ``synaptic'' connection to every intermediary node, and each intermediary node has a similar connection to every output node.
A typical run to learn the six  simple input-output patterns (i.e., maps) for a network of 6 input neurons, 300 middle neurons and 6 output neurons, is shown in (b)-(d). As noted in (b), the error eventually reaches zero, after which learning of a new pattern (c) is attempted.  The list of maps is shown in (d).
Re-drawn from~\cite{bak1}.
} 
\end{figure}

The toy model considered an arbitrary network of nodes connected by weights.
Although almost any network topology can be used, for the sake of description let us discuss the simplest version of a three feedforward layered network (see Fig.~\ref{fig1}-(a)).
To describe the working principle, let us suppose that we wish to train the network to learn an arbitrary map. To be precise, a map is an association from each input neuron to an output neuron. For instance, the identity map is one where each input neuron is associated to the corresponding output neuron (the first input neuron is associated to the first output neuron, the second input neuron is associated to the second output neuron, and the same relation holds for all other input neurons), and a random map is one where one output neuron is chosen randomly for each input neuron. 

 The learning algorithm needs to modify the network's weights in such a way that a given input neuron connects to the prescribed output neuron.
 
The entire dynamical process goes as follows:

\begin{enumerate} 
\item Activate an input neuron $i$ chosen randomly from the set established by the task to learn.
\item Activate the neuron $j_m$ in the middle layer connected with the input neuron $i$ with the {\it largest $w(j,i)$}.
\item Activate the output neuron $k_m$ with the {\it largest $w(k,j_m)$}.
\item If the output $k$ happens to be the desired one, nothing is done.
\item \label{deltastep} Otherwise, that is if the output is not correct, $w(k_m,j_m)$ and $w(j_m,i)$ are both reduced (depressed) by an amount $\delta$.
\item Go back to 1.
Another input neuron from the task set is randomly chosen and the process is repeated.
\end{enumerate}

The process involves a principle of extremal dynamics (here simplified by choosing the strongest weights) followed --in case of incorrect output-- by negative feedback (i.e., the step 5 of adaptation).
The only parameter of the model is $\delta$, but, at least in the numerical simulations, it is not crucial at all, because its only role is to prevent the same path from the input to the undesired output to be selected more than once.
Numerical explorations with this simple model showed that step \ref{deltastep} above can be modified in many different ways (including choosing random values) without serious consequences, as long as it makes less probable the persistence of ``wrong paths''. Initial values of $w(i,j)$ and $w(j,k)$ are not relevant either \cite{bak1}.

Notice that the toy model omits to consider the neuronal dynamics:  it is not necessary to introduce spikes whose only role would be
to propagate activity across the network. Since propagation occurs most often (statistically speaking) across the strongest synapses, the toy model omits including spikes and directly selects the strongest paths, as done in the steps 2 \& 3 of the algorithm.

In Fig.~\ref{fig1}-(b) and (c) we reproduce the results of a typical simulation in which few simple  maps (indicated on  Fig.~\ref{fig1}-(d)) are successively learned by the model (see details in~\cite{bak1}). 
The error in learning map ``a'', computed as the squared distance between the actual output and the desired one, is seen to fluctuate until eventually vanishing at time $\simeq 600$ (see Fig.~\ref{fig1}-(b)). After that, the network is given the task to learn map ``b'' (which is achieved at time $\simeq 800$), map ``c'' and so on. 
Interference between maps is expected for relatively small system size, since the same path can be chosen by chance for two different input-output maps.
As was discussed earlier in Ref.~\cite{bak1,bak2}, a system trained under these premises is robust with respect to noise, in the sense that depression of synaptic weights will self-adjust proceeding to correct the errors, until eventually achieving the desired  outputs.
Another interesting property of this set-up is that the learning time goes down with the size of the middle layer, a fact that is easily understood since the learning process implies to find and keep the strongest paths between the input and the desired nodes in the output layer.
This and other scaling relations can be found in Refs.~\cite{bak1,bak2,joe1,joe2,brigman}.

 \subsection{Memristor model}
 
 Now we turn to discuss how to implement the toy model just described on a network of memristors.
Memristive devices are a family of two-terminal devices whose resistance evolves according to the bias and currents they experience~\cite{strukov}.
In analogy to long term potentiation and depression taking place in neuronal synapses, memristor resistances can be increased or decreased through the application of relatively high voltage differences or currents.

{In this article, we will consider  Voltage Controlled Memristors with threshold \cite{MemEqs}, 
 whose resistance $R$, can take any value between  $R_{\text{min}}$ and  $R_{\text{max}}$.
When a voltage difference $V$ is applied, a current $I$ passes through the memristor, and the value of 
$R$ may change, depending on $V$ and $R$ values.}

\begin{figure}  
\vspace{-2.5cm}
\includegraphics [width=.95\linewidth]{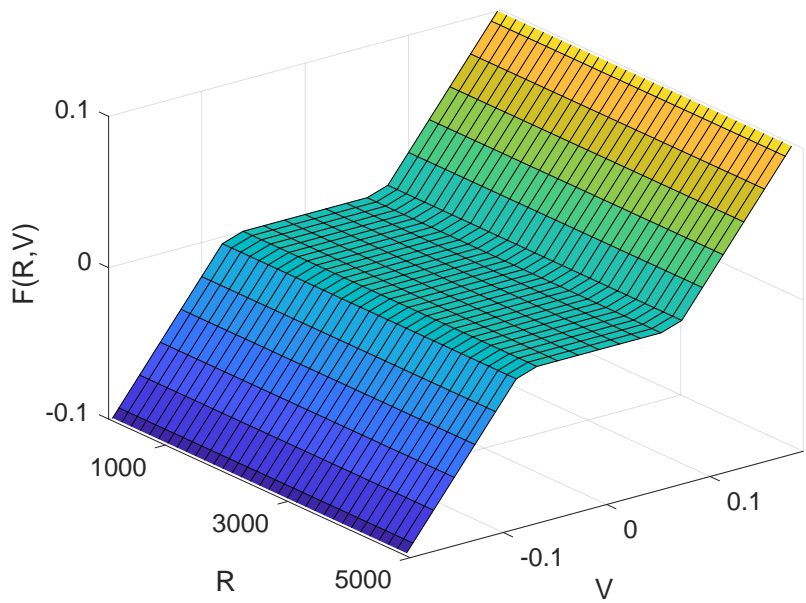}
\vspace{-3cm}
\caption{{Memristor behavior as a function of resistance and applied voltage.
The resistance of the memristor does not change unless the absolute value of the applied voltage is greater than $V_\updownarrow$.
Here we used $V_\updownarrow =0.075$, $R_\text{min}=75$, $R_\text{max}=5000$, $\beta=0.9$.}}
\label{FIG1SM}

\end{figure}

 {The memristor equations can be written as:
\begin{align} 
I &= V/R\\
\frac{\partial R}{\partial t} &= -F(R,V), \label{MemriEQV} 
\end{align}}

where the function $F$ describes the behavior of the memristor.
 Individual memristor dynamics is fully described by previous equations plus a definition of the function $F$, which describes how the characteristics of  the memristor change upon applied voltage differences. For that, we have used the bipolar memristive system with threshold (BMS), which is described  in detail in  Section 3.2 of \cite{MemEqs} (a different memristor model, known as Boundary Condition Memristor, BCM, \cite{PaperBCM,Comparison}, is considered in the Appendix).  According to that reference, we write:

\begin{equation}
F(R,V) = \begin{cases}
\beta (V + V_\updownarrow) & \text{if $V < -V_\updownarrow$, $R<R_{\text{max}} $ } \\
0 &\text{if $|V| <V_\updownarrow$} \\
\beta (V - V_\updownarrow) & \text{if $V > V_\updownarrow$, $R>R_{\text{min}}$ } 
\end{cases} \label{CasesV} 
\end{equation}

where $\beta>0$ is the rate at which the resistance increases or decreases when a large enough voltage difference is applied. The function $F(R,V)$ 
 is illustrated in  Figure~\ref{FIG1SM}. From Eqs. 1-3, we find that a relatively large positive voltage difference tends to decrease resistance, while a negative voltage difference tends to increase resistance on the memristor ($R$ is not modified if the absolute value of the voltage does not exceed the threshold $V_\updownarrow$).

\begin{figure}  
\centerline{
\includegraphics [width = 0.9 \linewidth] {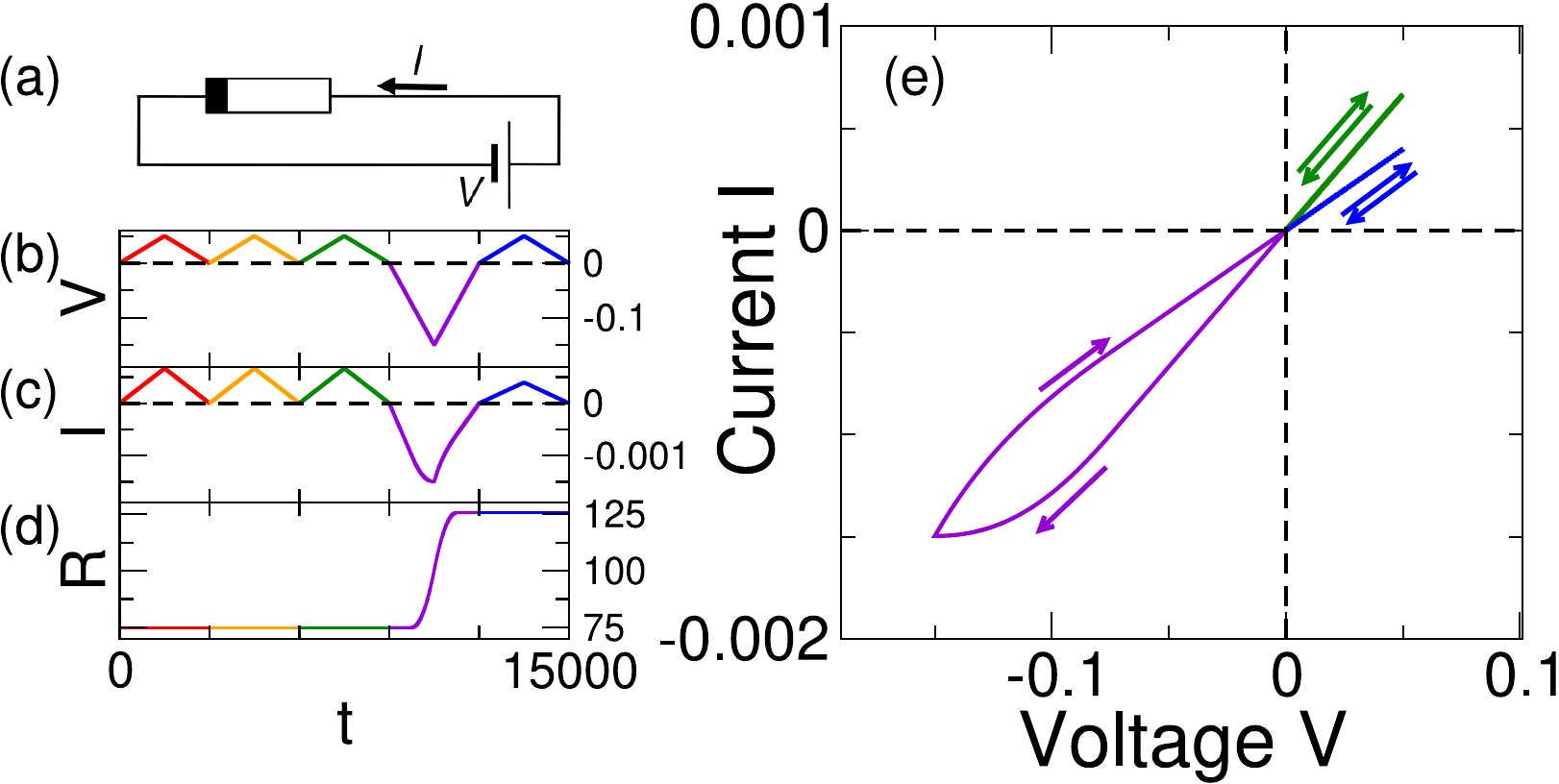} }
\caption{Behavior of a voltage controlled memristor connected to the circuit shown in  (a).
Typical changes in the properties of the memristor as a function of a time dependent voltage $V$  (b), 
the resulting current $I$ and the instantaneous resistance $R$ are shown in (c) and (d), respectively.
The same data is presented in panel as an I-V curve (e).
Notice that relatively small voltage excursions (i.e., upwards triangular sweeps) do not change the device resistance, while relatively large voltage excursions does it, resulting in the typical hysteresis loop.
Results in panels  (b)-(e) have been colored to ease the interpretation.}
\label{Fig2} 
\end{figure}

In Figure \ref{Fig2}, we show an example of the typical changes exhibited by voltage controlled memristor when subjected to  voltage sources of different amplitudes.
The circuit and the sign convention are depicted in Fig.~\ref{Fig2}-(a), while the voltage, the current across the device and the memristor resistance are shown in Figs.~\ref{Fig2} (b)-(d).
The voltage source applies three triangular shaped low \emph{positive}  voltage pulses, which do not change memristor's resistance, followed by a high, \emph{negative}, voltage excursion, which results on an increase of the memristor's resistance.
The final triangular low voltage pulse shows that the resulting resistance increase is permanent.
This property will be used here to modify the network input/output paths, as explained in the following paragraphs \cite{units}.

 \subsection{Memristor Network}

\begin{figure}  
\centerline{
\includegraphics [width = .6\linewidth] {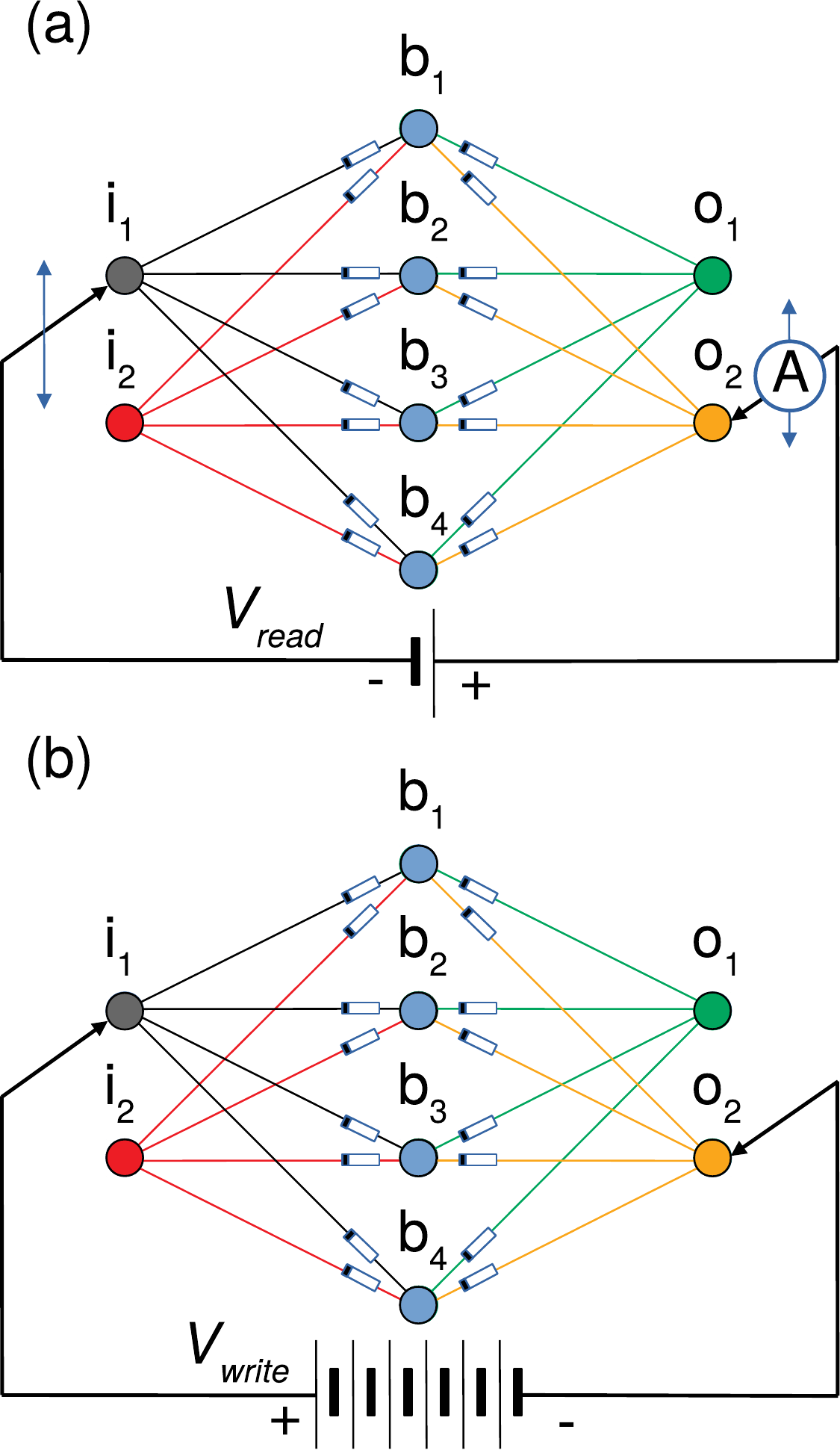} }
\caption{Sketch of the learning algorithm for a network with $N_\text{in}=2$, $N_\text{bulk}=4$ and $N_\text{out}=2$.
In the Reading step (a)  a relatively small $V_\text{read}$ voltage is applied and the current at each output node is measured by the ammeter.
The output node with the largest current is defined as the output.
 If that output is not the desired one, in the Correction step (b) 
  a relatively large $V_\text{write}$ voltage  is applied to alter the resistance of the memristor path.
The cycle is repeated until the desired map   is learned.}
\label{fig2018} 
\end{figure}

The results presented here correspond to numerical simulations of a three layer network of memristors \cite{code}, with $N_\text{in}$ input nodes, $N_\text{bulk}$ bulk nodes, and $N_\text{out}$ output nodes.   Pairs of nodes from successive layers are connected through bipolar memristive system with threshold (BMS) of Ref.~\cite{MemEqs} (see also Ref.~\cite{MemEqs2}). Notice that this three-layer network  is equivalent to a  
memristor crossbar  array \cite{cross} with $N_\text{in}$ input nodes and $N_\text{bulk}$ output nodes, connected to a second 
memristor crossbar array, of $N_\text{bulk}$ inputs and $N_\text{out}$ outputs.
The training algorithm uses an ammeter and a voltage source with two possible values: $V_\text{read}$ (read voltage) and $V_\text{write}$ (punishment or correction voltage). Memristor polarity is set in such way that a negative $V_\text{write}$  tends to increase their resistance (see Fig.~\ref{Fig2}-(a) and Fig.~\ref{fig2018}-(b)).
The network and control resources are as sketched in Fig.~\ref{fig2018}.

  {We considered a relatively small (quenched) variability in the parameters of the device: 
 the parameters for each memristor were randomly chosen from a uniform distribution with   $0.8<\beta<1$, $0.05<V_\updownarrow<0.1$, $50<R_{\text{min}}<100$, and $R_{\text{max}}=5000$.
Initial condition was set to $R=R_{\text{min}}$.
The reading step lasted $1$ time step using a voltage value $V_\text{read}=0.0001$.
The correction step lasted 5 time steps using a voltage $V_\text{write}={-}0.2$.}

 {Voltage values were chosen such that  $|V_\text{read}| \ll V_\updownarrow < {|}V_\text{write}/2 {|} $.
In this way the memristor properties do not change on the reading step, and only few memristors change its resistance during the correction step of the algorithm.}

{\subsection{Learning Algorithm}}

Here we discuss the implementation defining a simple input/output association task, similar to the one already discussed in the previous section for the case of the toy model of Fig.~\ref{fig1}: for each input node $i_n$, we ask the memristor network to learn a randomly chosen output node. 

The proposed training involves the following sequence at each training step:

\begin{enumerate} 
\item \label{choosestep}Randomly choose an input node ($i_n$).
\item \label{readstep} Read the current flowing through all the output nodes.
 Herein this is done by setting $i_n$ voltage to $V_\text{read}$, and moving the ammeter tip (which closes the circuit) to each output node sequentially ($o_1$, $o_2$, etc), while measuring the current (see Fig.~\ref{fig2018}-(a)).
\item Determine the output node with the maximum current.
\begin{enumerate}
 \item If the node with maximum current is the desired one, do nothing.
 \item \label{punishstep}Otherwise, that is if the output maximum current is not at the desired node, apply $V_\text{write}$ (see Fig.~\ref{fig2018}-(b)), and go back to point 2.
\end{enumerate}
\item Go back to point 1.

\end{enumerate} 

The  value of $V_\text{read}$ needs to be small (such that it does not change the values of the resistances in the network).
$V_\text{write}$ is large and with inverted polarity, hence inducing an increase of the resistances of the network.
This is the only crucial factor, to ensure that the reading (in step \ref{readstep}) is not modifying the network conductances and conversely that the correction (in step \ref{punishstep}) decreases the likelihood of having large currents in the undesired paths.
It is evident that the memristor learning algorithm preserves the same spirit of the earlier work: to punish wrong paths by increasing the resistance of the involved memristors.
We applied step~\ref{punishstep} a maximum of $n_\text{max} = 80$ times on each training step \cite{nmax}.

To collect the statistics presented here, at the end of each  training step we calculate the learning error as follows. For each input node, we find the largest output  (we apply $V_{read}$ among that input node and all output nodes sequentially, and take the node through which current flow is largest).
 We define the error as the Hamming distance from the vector of largest outputs and the desired map.
If the error is null, the network has learned. Otherwise, a new training step is performed.
In some cases, after the network has learned, we will consider to train it with a different map. The pseudocode for this algorithm is shown in the Appendix.

\begin{figure}  
\includegraphics [width=.9\linewidth]{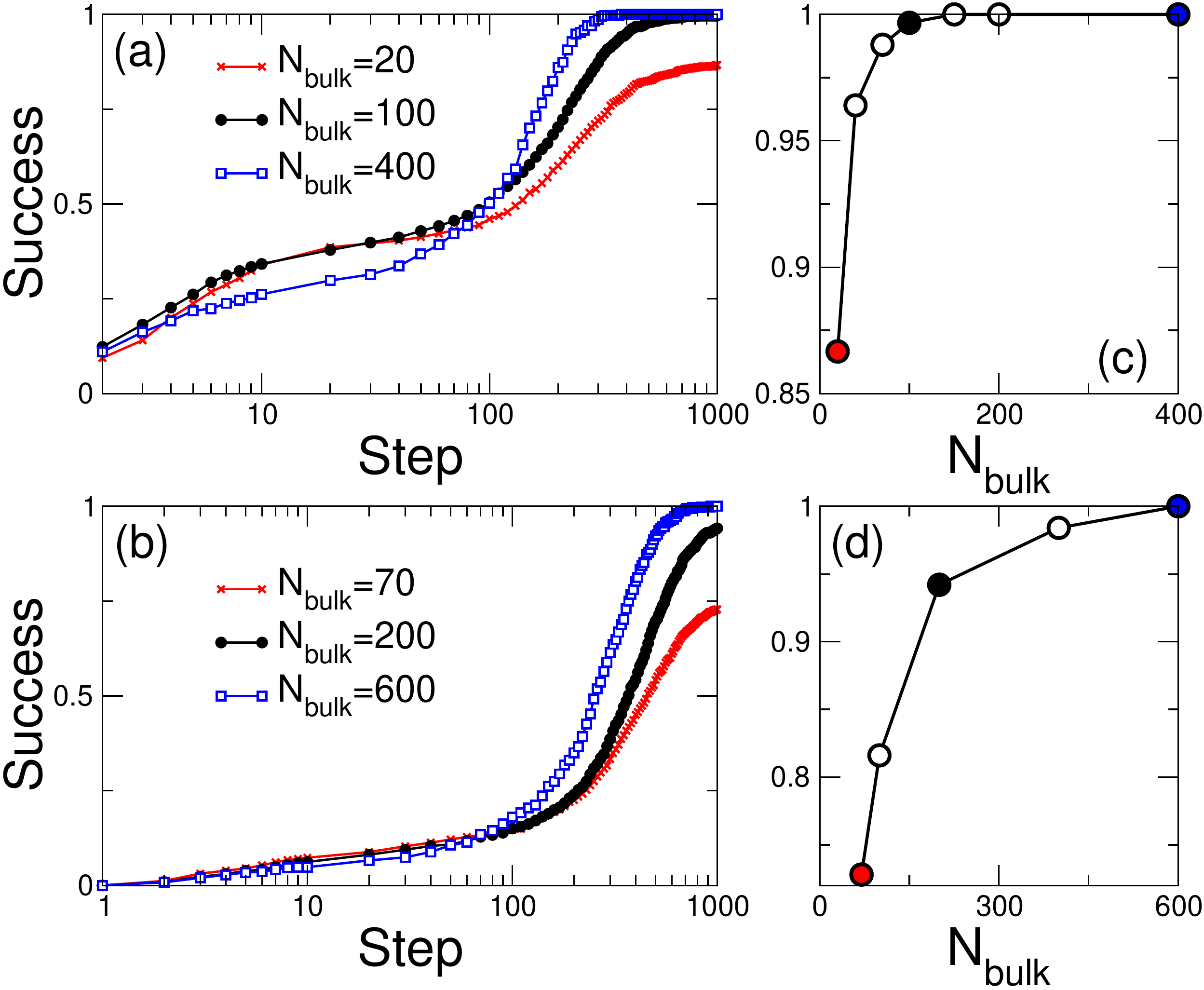}
\caption{
Learning performance as a function of the middle layer size.
Results show the success as the fraction of networks that learn a random input/output association map after a given number of correction steps for $N_\text{in}=N_\text{out}=3$, $N_\text{bulk} = 20$, $100$ and $400$  (a), and for  $N_\text{in}=N_\text{out}=4$, $N_\text{bulk} = 70$, $200$ and $600$ (b).
Success at 1000 steps, as a function of  $N_\text{bulk}$ for $N_\text{in}=N_\text{out}=3$ 
(c) and for $N_\text{in}=N_\text{out}=4$ (d) are also shown. 
In (c) and (d) the coloured symbols correspond to the results in (a) and (b) obtained with the respective $N_\text{bulk}$ values.
All results are averages over at least 500 network realisations, yielding values of standard errors smaller than the symbols size ($SEM \sim 0.025$).
}
\label{Figure5}

\end{figure}

 {\section{Results}}
 \label{Results}
Now we proceed to describe the parametric behavior of the algorithm just explained in the previous paragraphs. 
First we explored the dependence of the learning time on the size of the middle layer $N_\text{bulk}$ for random maps.
As discussed, larger values of $N_\text{bulk}$ in the neuron network model of Ref.~\cite{bak1,bak2} provided more paths to the correct output, which leads to shorter learning time.
We found very similar performance for the memristive network as shown in the results of Fig.~\ref{Figure5} where success (the fraction of networks that have learned a map) is shown as a function of the training step and $N_{bulk}$ for a three layer networks with $N_\text{in}=N_\text{out}=3$, and $N_\text{in}=N_\text{out}=4$, and several values of $N_\text{bulk}$.
In panels (a) and (b), success 
as a function of the step number improves with larger $N_\text{bulk}$.
This is also apparent when we plot the fraction of networks that have learned at (or before) correction step 1000 (see panels (c) and (d)) showing that performance is an increasing function of the middle layer size, $N_\text{bulk}$. It can also be noticed, from Fig.~\ref{Figure5}-(c) and (d), that Success in equal to $1$ for long enough 
$N_\text{bulk}$, which means that any map can be learned on those networks. 
{Similar numerical simulation results, for  a different memristor model \cite{PaperBCM,Comparison}, with parameters fixed to represent the behavior of the first reported memristor \cite{strukov,Pickett}, are shown in the Appendix.  This similarity  suggests that the performance of the learning algorithm presented here does not depend strongly on the details of the memristors used. } 

\begin{figure}  
\includegraphics [width=.9\linewidth]{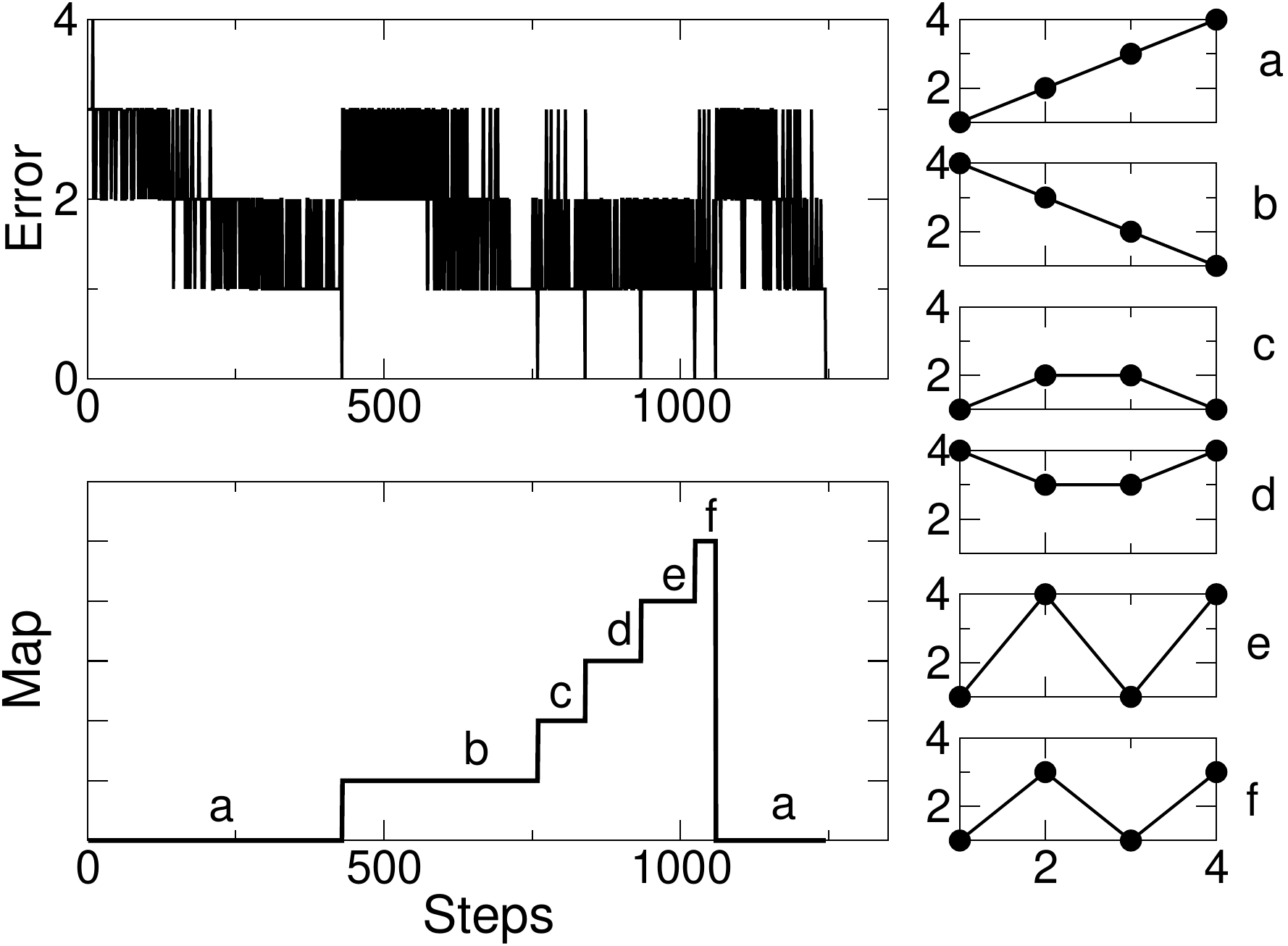}
\caption{Typical evolution of the training of a network learning seven successive maps.
Upper main panel shows the error (Hamming distance from the desired to the current output) as a function of training steps.
The input/output maps, labeled \textbf{a} to \textbf{f}, are depicted in the right column and presented sequentially as indicated in the lower main panel.
Network with $N_\text{in}=N_\text{out}=4$, $N_\text{bulk} = 200$, other parameters same as in Fig.~\ref{Figure5}.
} \label{Figure6}
\end{figure}

The implementation of the training algorithm shows also that the memristor network learns a series of maps in a way similar to that exhibited by the earlier neuronal model~\cite{bak1,bak2}.
This can be seen in the example of Fig.~\ref{Figure6}, which shows the evolution of the network with $N_\text{in}=4$, $N_\text{bulk}=200$, $N_\text{out}=4$.
The network was trained in one of the labeled maps until eventually the error is zero, at this point it starts being trained on a different map and so on.
Notice the resemblance with the results in Fig.~\ref{fig1}, which suggests that the training strategy proposed here is capturing the essence of the learning algorithm of Ref.~\cite{bak1,bak2}.

\begin{figure}  
\centerline{ \includegraphics [width = .55\linewidth] {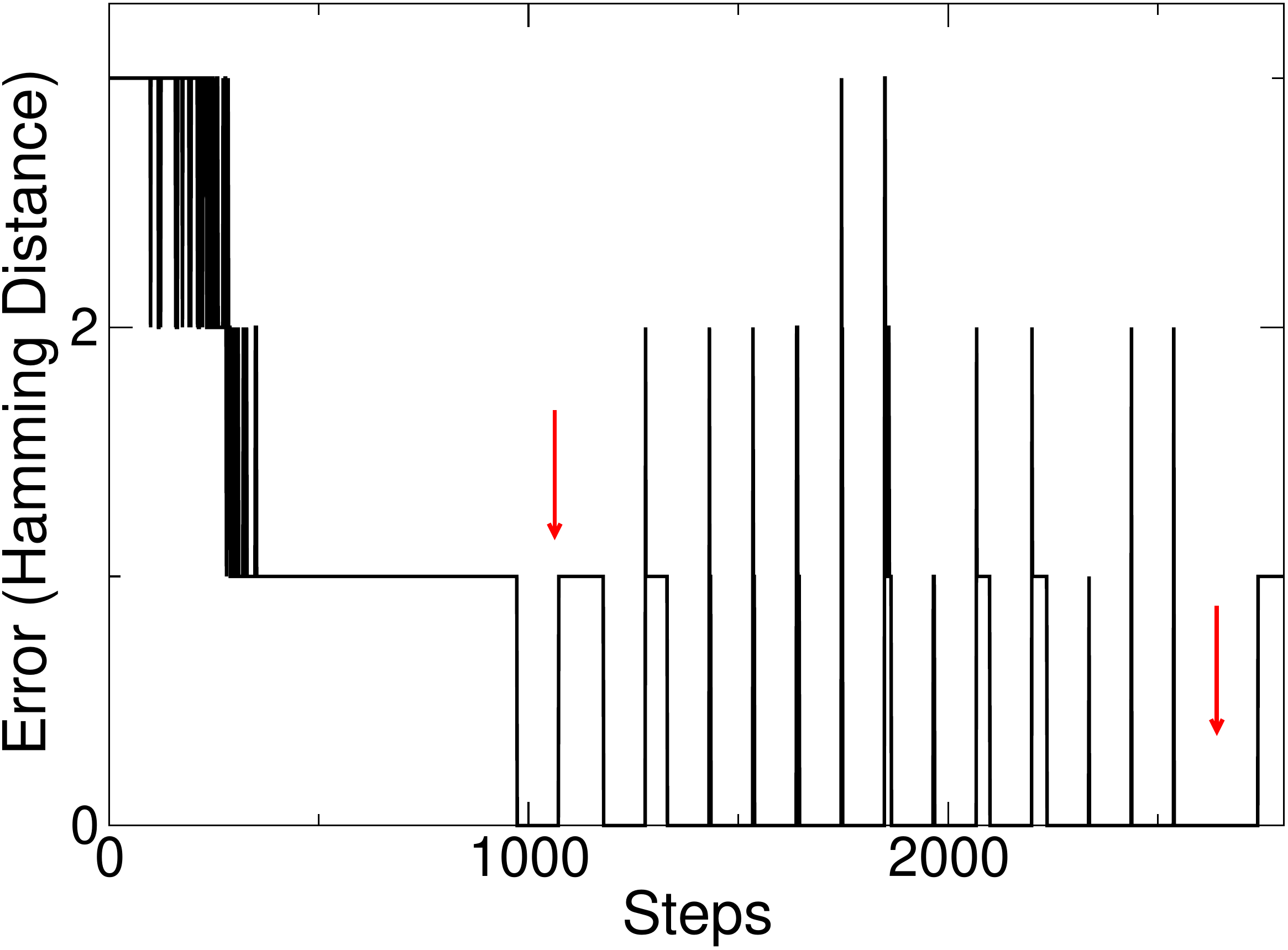} }
\caption{Example of the network recovery after a single perturbation.
We plot the error (i.e., the Hamming distance from the desired to the current output) as a function of the training steps for a network which is learning the identity map.
100 steps after the map is learnt, $10\%$ randomly chosen memristors are perturbed by increasing their resistances by $5\%$.
 The first downward arrow indicates the first perturbation, while the second downward arrow denotes the 14th perturbation (which did not increase the network error).
Network with $N_\text{in}=N_\text{out}=4$, $N_\text{bulk} = 200$, other parameters same as in Fig.~\ref{Figure5}.}  \label{Figure7}
\end{figure}

Another distinctive property of the proposed training strategy is the fact that the memristive networks are robust to perturbations of the device properties.
This alterations can be seen, for instance, as changes in resistance, which in real networks can be due to volatility, defects, etc.
As an example, we plot in Fig.~\ref{Figure7} the evolution of a network ($N_\text{in}=N_\text{out}=4$, $N_\text{bulk} = 200$), which after learning the identity map, %
is periodically perturbed.
It can be seen that after each perturbation, the network recovers to null error learning the map in a few additional steps.
This ability is not surprising given the fact that the network proceeds with the perturbation in the same way than during the usual learning process.

To gain insight on the dynamics of the memristor network resistances during learning of successive maps a $N_\text{in}=N_\text{out}=3$, $N_\text{bulk}=400$ network was trained to learn all the possible $N_\text{out}^{N_\text{in}}=27$ maps.
After all maps were learned, the network location of each memristors were randomly shuffled (preserving their resistance values).
After that, the network was re-trained to learn the same $N_\text{out}^{N_\text{in}}$ maps.

\begin{figure}  
\centerline{ \includegraphics [width = .9 \linewidth] {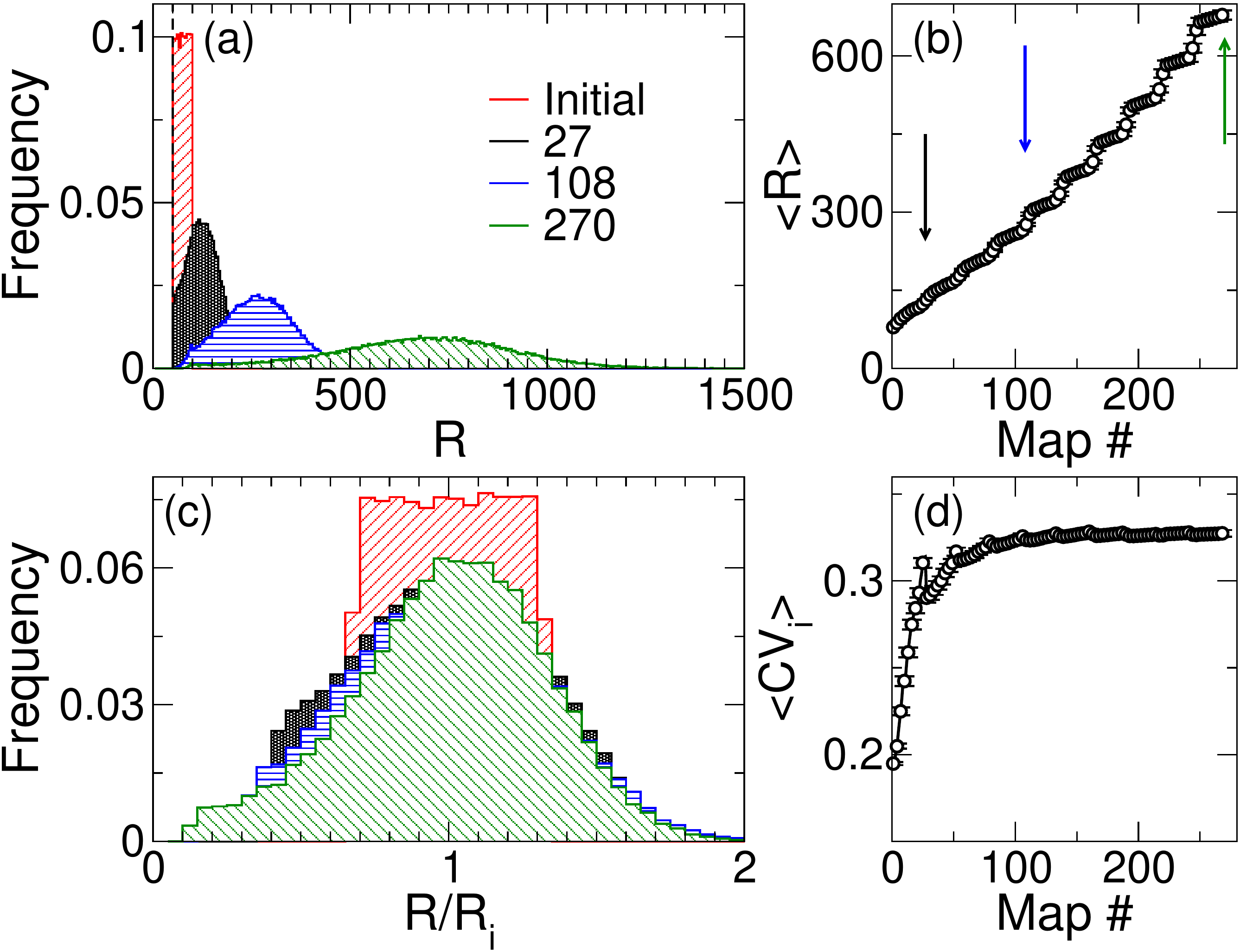}} 
\caption{Evolution of the distribution of resistance values after repeated re-learning of the same set of maps.
Resistance histogram at the beginning of the simulation, after learning all possible maps once (labeled 27) after shuffling and relearning all maps four times (labeled 108) and after shuffling and re-learning all maps ten times (labeled 270) are shown in (a).
The dashed vertical line shows the minimum possible value $R_{\text{min}}=50$.
Average resistance as a function of number of learned maps is shown in (b).
Arrows indicate the maps used for the histograms in  (a).
 The same data as in (a) after normalising each memristor's resistance by the average value of the whole network is shown in (c).
 Coefficient of variation of memristor's resistance, as a function of the number of learned map is shown in (d).
 Results are calculated for a single network and then averaged over 25 network realisations.
Bin-size $\Delta R=5$ for histogram in (a)  and $\Delta R=0.05 R_i $ in (b).
Network with $N_\text{in}=N_\text{out}=3$, $N_\text{bulk} = 400$, other parameters values as in Fig. \ref{Figure5}.} 
 \label{Figure8}
\end{figure}

In Fig.~\ref{Figure8}-(a), we show the histogram of resistances, at the beginning of the simulation, after learning all possible maps once (labeled 27) after shuffling and relearning all maps four times (labeled 108), and after shuffling and re-learning all maps ten times (labeled 270).
In panel b, we show the evolution of the average resistance, as a function of the number of learned maps.
As expected from the nature of the training algorithm, resistances can only grow when performing correction steps.
Moreover, from Fig.~\ref{Figure8}-(a), the range of resistance values increases with the number of steps.
We remove the effect of growing resistance by normalizing with the average resistance of the $i$th network, $R_i$.
In Panel (c) we plot a histogram of the normalized resistances, where each resistance value was divided by the average resistance of the network.
Finally, in Panel (d), we show the coefficient of variation, $<CV>$, of a single network, computed as the standard deviation divided by the mean value of all resistances in the network.
After repeated learning the distribution tends to a Gaussian distribution, approaching a $<CV> \sim 1/3$.

{\section{Discussion}}


Several learning strategies for memristor networks have been recently proposed \cite{Soundry,Hasan,4memr,reservoir}. 

For  problems with time dependent  inputs, one strategy consists on using \emph{reservoir computing} \cite{CarbajalRC}, where  the inputs are connected to a reservoir network composed of  interconnected time dependent elements (whose properties  continue evolving  \emph{after} a stimulus is applied), such as \emph{volatile} memristors. The response of the network is then classified through an output layer, that needs to be trained.  Reservoir computing has also been applied to time independent inputs that are re-encoded as time-dependent ones \cite{reservoir}. While this kind of networks differ from those studied here, an error penalization learning mechanism, such as the one studied here, may be useful for training the output layer of this scheme.

For time-independent inputs,  several learning strategies considering  \emph{non-volatile} memristors (as done here), have also been proposed. Most of them are inspired in \emph{Machine learning} algorithms, such as \emph{Gradient descent} training (see e.g. \cite{Soundry,Hasan}), where the value of each resistance needs to be known and modified at each correction step. As an alternative,  a \emph{random weight  change  algorithm}, which does not require to know the precise values of all elements, has recently been proposed  for training a network  where each synapse is composed of four memristors, and several transistors \cite{4memr}.

Similarly, here we have described a training procedure allowing a {simple} network of memristors to learn any arbitrary input/output association map.
This is achieved without any detailed information of the inner structure of the layers. 
{The method is inspired by a learning  algorithm  proposed about 20 years ago \cite{bak1,bak2}, however, there are some differences among them.}

{First, in the original model, the active neurons  transmitted their activity to just one neuron on the following layer, while here, the current passing through each node is determined by Kirchoff's laws. Probably, the inclusion of diodes or transistors (such as in Refs. \cite{reservoir,4memr}) may  generate  dynamics which are closer to the original proposal. The study of this possibilities is an interesting avenue for future research.}

{Second, in the correction step for the toy model, only the (two) intervening connections are punished.  Here, instead, we apply $V_{write}$ over the input and output nodes, in such way that the correction voltage difference (and consequently the correction) over each memristor in the network is  proportional to the voltage involved in the wrong answer. In this way, the correction is performed without the need to measure or to  have control over  the middle layer:}
the learning method  does not need to control individual memristors and  requires only access to read and/or perturb two nodes
from the (arbitrarily large) networks at any given time (consider, as an example, the network of Fig.~\ref{Figure8}, which has 2400 memristors, and only  3 input nodes plus 3 output nodes need to be considered in the present learning method).

Despite the extreme simplicity of the approach, it is demonstrated that an iterative reading of the current flowing between two points of the network, and its eventual perturbation, can modify the overall network connectivity until arriving to one of the possible solutions.
By design, the learning is robust against different types of perturbations, including differences and fluctuations in the memristors parameters, noise and defects as well as distribution of polarities (see the Appendix).
Although the present results are limited to numerical simulations, in case of being implemented in hardware, the method is easily scalable to arbitrarily large network sizes.

Notice that the memristor network is able to learn despite lacking an important feature of its biological neural contra-part: the spikes. 
That is not a limitation, because as commented in the introduction, in this kind of  networks, the only role for the spikes would be to propagate the information from the input node/s through the network to some output node/s.
Instead, in the present implementation, this is also achieved, but by the current flow from an external battery.
Thus, knowing which input node is connected, the current flowing through the network will be reflected on the value of the output current at a given node.
In this context, our approach is a simple solution which does not require at all of the implementation of any spiking mechanism.
Obviously, the absence of additional electronics to implement neurons becomes very relevant when considering a hardware implementation of this concept.
Concerning hardware, the proposed learning method benefits from variability in the network properties, therefore not requiring of precise control on the memristor parameter during manufacturing.

{Several analytical results have been provided for the original model \cite{joe1,joe2}.  For instance, it has been shown that a geometric transition from no-learning to learning takes place at $N_{Bulk}= N_{in} \times N_{out}$. While  our numerical simulation results show similarities with the original  learning algorithm, it would be  useful to verify whether and how these results are valid here. Also, extending   other analytical approaches, such as the work by Caravelli \emph{ et. al.} \cite{Caravelli2,CaravelliScience} to the system studied here, should give a deeper understanding on how learning in this memristor network takes place.  } 

In these notes we limited ourselves to the presentation of the most fundamental aspects of the results.
As a salient feature, we reported how a single network can learn many maps, and this causes an expansion of the distribution of their resistances, possibly due to the multiplicity of solutions to learn a single map.
Nonetheless some few caveats must be mentioned.
In the first place we consider adversarial situations for the algorithm, concerning some simple variations on the approach, where changes in the initial condition of memristor conductances, polarity as well as minor changes in the type of correction step are described. 
These studies are presented in the Appendix.
Second, we did not expand here on discussing the type of problems that the present approach can solve.
This issue would require of extensive numerical simulations and it seems to deserve being explored on a hardware implementation, since it will work tens of orders of magnitude faster than any of our current numerical simulations. Finally, we expect the approach  to be useful on a variety of  network topologies including  less ordered systems such as a random   network of nanowires \cite{nanowires} whose conductivity can be varied by applying a voltage difference among pairs of points in the network. The algorithm may also be useful on other structures \cite{dots} as long as a correction mechanism increasing resistance over not desired paths can be generated.
\\
\\
\section{Conclusions}

In summary, we have introduced an algorithm able to train a memristor network to learn arbitrary associations.
Robust results for its performance were demonstrated using numerical simulations of a network of voltage controlled memristive devices.
Given the design principles, the results suggest that its implementation in hardware would be straightforward, being scalable and requiring very little peripheral computation overhead.


\paragraph{Acknowledgments.}
 Work partially supported by 1U19NS107464-01 NIH BRAIN Initiative (USA) and CONICET (Argentina).


\newpage
\section*{Appendix}

\renewcommand{\thefigure}{A\arabic{figure}}

\setcounter{figure}{0}

\subsection{Miscellaneous observations} 
\begin{figure} [] 
\includegraphics [width=.85\linewidth]{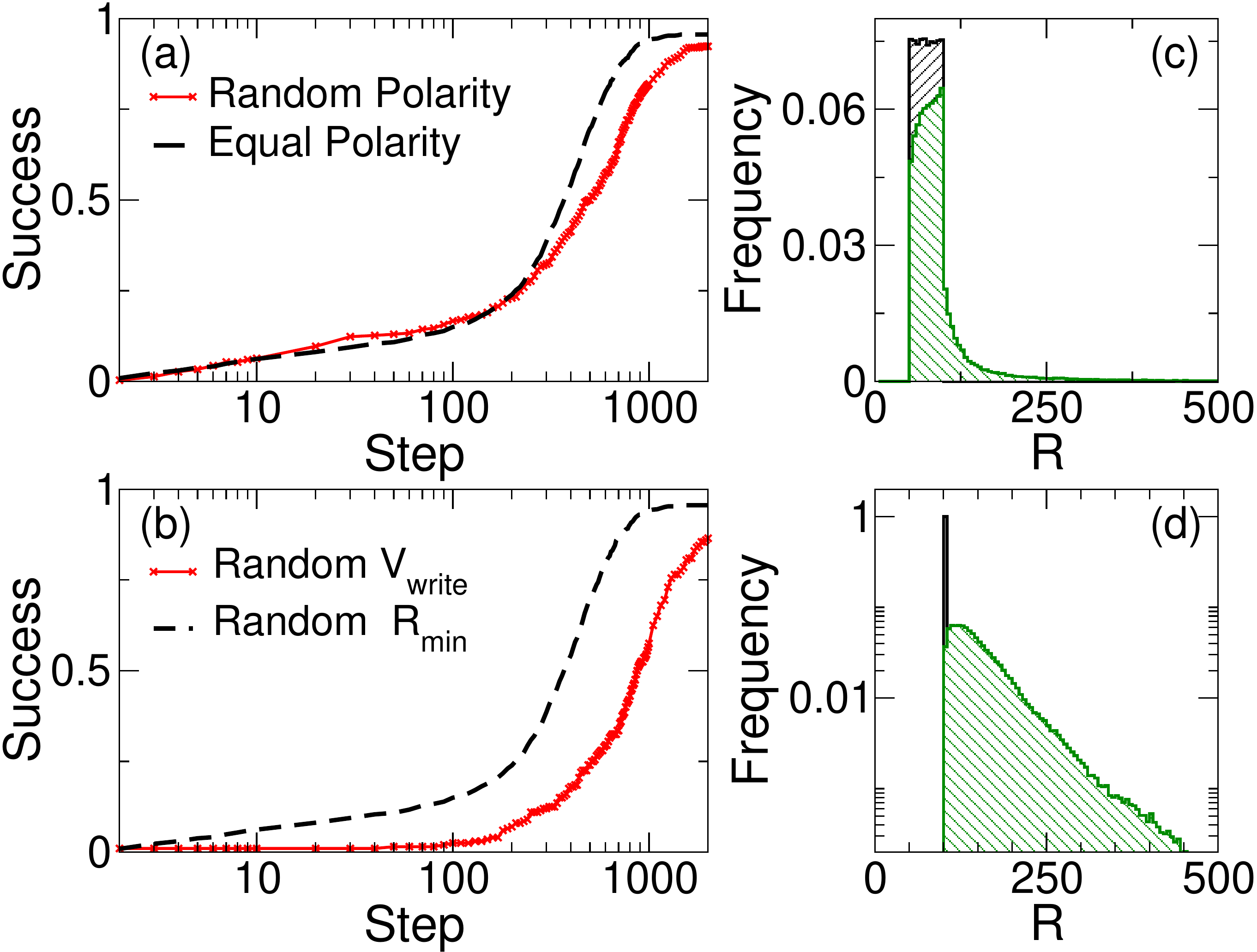}
\caption{Performance of the approach using {a} random distribution of memristor polarities or random $V_\text{write}$ correction values.
Success as a function of step number for networks where memristors polarities are chosen at random {is shown in panel (a)}.
 Success as a function of step number for equal initial conditions $R_\text{min}=100$, and correction voltages $V_\text{write}$ uniformly drawn from $0.15$ to $0.3$ {is shown in panel (b)}.
 Initial and final resistance density distribution for the case of using random polarity, and  random correction $V_\text{write}$ {are shown in (c) and (d) respectively}.
For comparison, in Panels {(a) and (b)} the dashed line reproduces the results presented in Fig.~4-(b) ($N_B=200$, black circles).  In (c) and (d), bin-width $\Delta R=5$. Results are averaged over at least 100 network realizations, using in all cases, $N_\text{in}=N_\text{out}=4$, $N_\text{bulk}=200$.
}
\label{FIG2SM}
\end{figure}

Some special cases are described here, including different initial conditions in the memristor parameters and variations on the implementation of the correction step, noting that all the results remain valid despite these changes. 
 Fig.~\ref{FIG2SM}-(a) shows the results of simulations where the polarity of the memristors were distributed randomly.
As a comparison we plotted (with dashed line) the data presented already in Fig.~4-(b)  corresponding to equal memristor polarity.
Panel (c) shows the changes in the $R$ distribution before and after the learning process.
 These results suggest that the distribution of polarities  produces only minor changes in the overall performance of the approach.

Then we explored how may  affect the performance the lack of variability on memristor' properties, by  starting the simulation with identical resistances ($R=100$) for all memristors, and using random $V_\text{write}$ values in the correction steps (uniformly distributed from $0.15$ to $0.3$) while maintaining the other parameters $\beta$ and $V_\updownarrow$ randomly distributed.
Results are shown in { Fig.~\ref{FIG2SM}-(d)}.
It is apparent that the network learns approximately in the same manner as starting with random initial conditions for $R$, except that it takes additional steps to reach comparable success rates. Probably these additional correction steps are trivially related to the time needed to generate a minimal dispersion on the $R$ values, needed for the approach to work. Thus, the manufacturing variability of the memristors properties expected on an experimental setup will not be disadvantageous.
In Fig.~\ref{FIG2SM}-(d) the initial (i.e, $R=100$) and final distribution of resistances for this case are shown, showing the resulting broad $R$ distribution, after the map is learned.
\\
\\
\\
\subsection{Alternative  memristor model}
We have reproduced some results in the main text using a different memristor model, known as Boundary Condition Memristor (BCM). This  model was proposed in Ref.  \cite{PaperBCM}.  In Ref. \cite{Comparison}, it was shown that BCM model reproduces  the current-voltage characteristics of Pickett's model \cite{Pickett} (which explains the dynamics of the first reported memristor, based on $TiO_2$ nanofilms \cite{strukov})   closer than other alternative descriptions, when the parameters are chosen adequately. 

The equations for a single BCM memristor can be written as a function of a state parameter $\omega$ as:

\begin{align} 
I &= V/R \tag{A1}\\
R &=R_{max}-{\omega \over D} (R_{max}-R_{min})\tag{A2}\\
\frac{\partial \omega}{\partial t} &=  {\mu R_{min} \over D} I f_B(\omega,V), \label{MemriBCM}\tag{A3}
\end{align}

{where $D=10 nm$ is the assumed width of the memristor  ($0\leq {\omega\over D} \leq 1$), and $f_B$ may have three different values $(a>0,b>a,0)$ depending on 4 nonnegative constants:
$v_{t0}$,$v_{t1}$,$v_{th0}$, $v_{th1}$, and the value of $\omega$ and $V$: }

{For $0< {\omega\over D} < 1$,}

\begin{equation}
{f_B(\omega,V) = \begin{cases}
a & \text{if $-v_{t1} \leq V \leq v_{t0}$  } \\
b &\text{otherwise.} 
\end{cases} }\label{CasesV1}\tag{A4}
\end{equation}

{For ${\omega\over D} =1$ (minimum resistance),} 

\begin{equation}
{f_B(\omega,V) = \begin{cases}
b & \text{if $V <-v_{th1}$  } \\
0 &\text{otherwise.} 
\end{cases} }\label{CasesV2}\tag{A5}
\end{equation}

{For ${\omega\over D} =0$ (maximum resistance), }

\begin{equation}
{f_B(\omega,V) = \begin{cases}
b & \text{if $V >v_{th0}$  } \\
0 &\text{otherwise.} 
\end{cases} }\label{CasesV3}\tag{A6}
\end{equation}

{The parameters which closer reproduced Pickett's result, having $D=10nm$, $\mu= 10^{-16} m^2 V^{-1} s^{-1} $, $R_{min}=10^3  \Omega$ and $R_{max}=10^4 \Omega$ fixed, were: $a=0.1494$, $b=1.6182$,  $v_{t0}=0.915\, V$, $v_{t1}=1.3048\, V$, $v_{th0}=4.7404\, V$, and $v_{th1}=2.4629\, V$, ($a$ and $b$ are unit-less), see Ref. \cite{PaperBCM}.}  

\begin{figure}  
\vspace{-2.5cm}
\includegraphics [width=.95\linewidth]{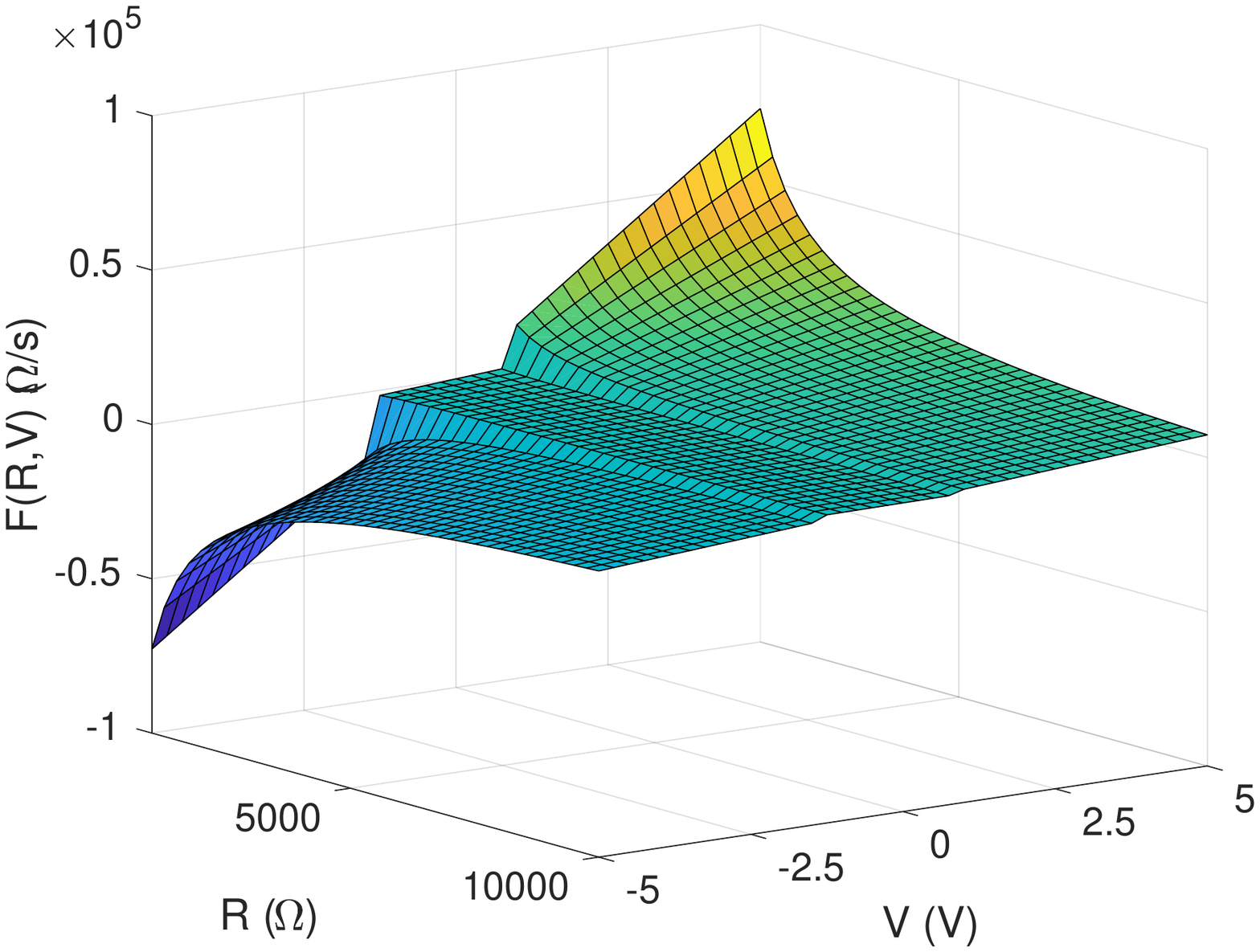}
\vspace{-2.5cm}
\caption{{BCM Memristor behavior as a function of resistance and applied voltage, using the parameters proposed in  Ref. \cite{PaperBCM} ($D=10nm$, $\mu= 10^{-16} m^2 V^{-1} s^{-1} $, $R_{min}=10^3  \Omega$, $R_{max}=10^4 \Omega$, $a=0.1494$, $b=1.6182$,  $v_{t0}=0.915\, V$, $v_{t1}=1.3048\, V$, $v_{th0}=4.7404\, V$, and $v_{th1}=2.4629\, V$)}.}
\label{FIGSM2}
\end{figure}

{Eqs. A2 and A3   can be rewritten in terms of Eq. \eqref{MemriEQV} setting $F(R,V)={\mu (R_{max}-R_{min}) R_{min}  \over D^2} {V\over R} f_B(\omega(R),V)$. The function $F(R,V)$ for BCM model is shown in Fig. \ref{FIGSM2}. Notice that the BCM model presents some differences with the model in the main text (BMS). The most important one is that now the correction function $F$ depends on the current, $I=V/R$. This means that, intuitively, in the first model, highest resistance values would tend to show larger voltage differences, and thus, to have stronger  corrections (for instance, when connected in series with another resistance), while here, lowest resistance values will tend to have higher corrections. Also, the model presents four  different voltage thresholds, and resistance evolution even with small voltages applied (for $a>0$). } 

\begin{figure}  
\includegraphics [width=.9\linewidth]{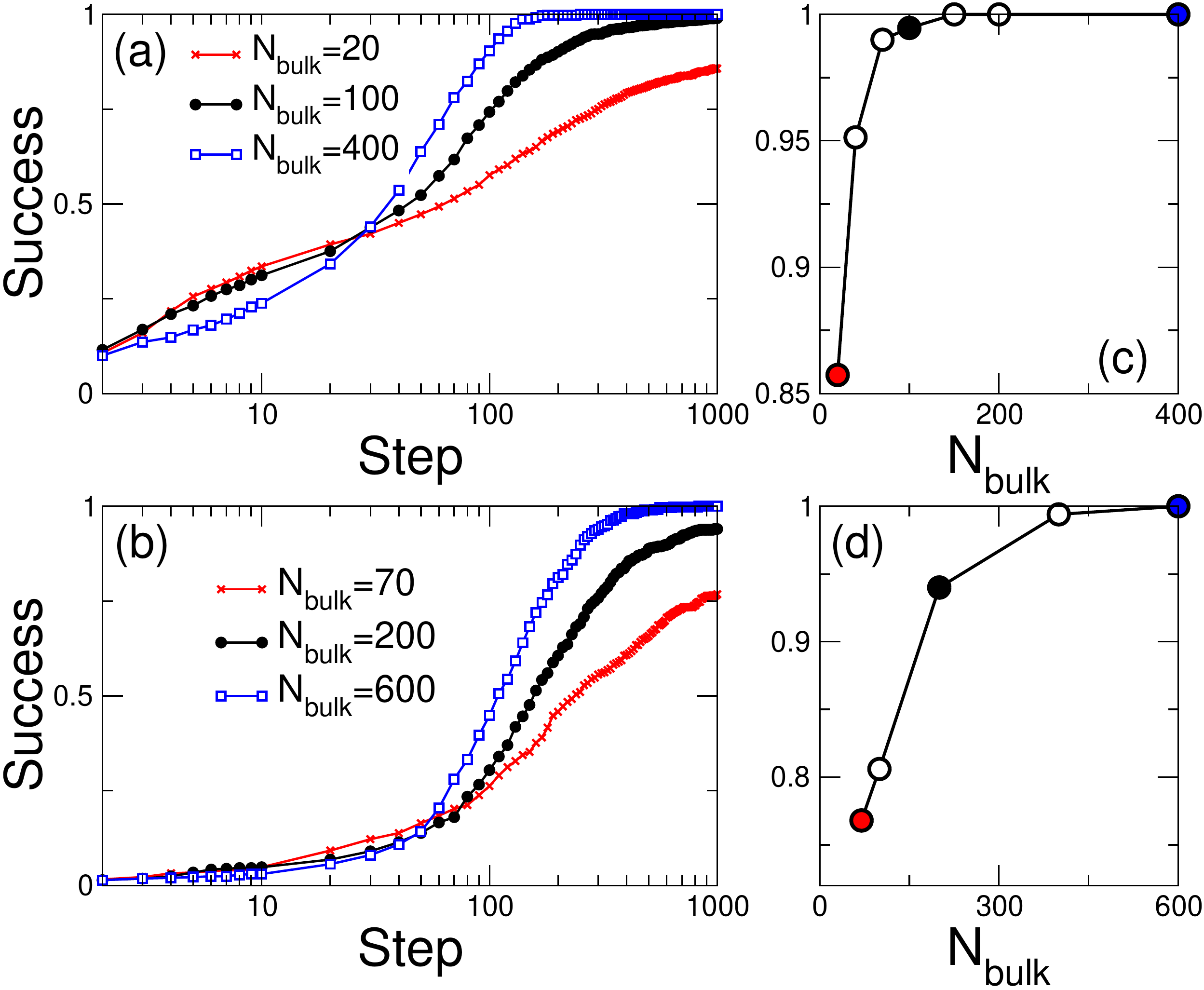}
\caption{
Learning performance as a function of the middle layer size, for BCM memristors.
Results show the success as the fraction of networks that learn a random input/output association map after a given number of correction steps for $N_\text{in}=N_\text{out}=3$, $N_\text{bulk} = 20$, $100$ and $400$  (a), and for  $N_\text{in}=N_\text{out}=4$, $N_\text{bulk} = 70$, $200$ and $600$ (b).
Success at 1000 steps, as a function of  $N_\text{bulk}$ for $N_\text{in}=N_\text{out}=3$ 
(c) and for $N_\text{in}=N_\text{out}=4$ (d) are also shown. 
In (c) and (d) the colored symbols correspond to the results in (a) and (b) obtained with the respective $N_\text{bulk}$ values.
}
\label{FIGSM3}

\end{figure}

We have reproduced the results shown in the main text, using BCM memristors whose evolution is given by Eqs \ref{MemriBCM}-\ref{CasesV3}, and the parameters listed above, except  $R_{min}$, which is randomly chosen between $500$ and $1000 \Omega$ for each memristor. The algorithm parameters are now $V_{read}=0.0001 V$, $V_{write}=-5 V$, and the time over which the correction voltage is applied is set to $\Delta t=1 ms$. Results similar to those presented in main text are shown in Fig. \ref{FIGSM3}. A detailed analysis of the  performance of this method as a function of the  values of memristor parameters  ($v_{t0}$,$v_{t1}$,$v_{th0}$, $v_{th1}$, $R_{min}$, $R_{max}$, $a$,$b$), including noisy parameter distribution, or the parameters of the learning algorithm ($V_{read/write}$, $\Delta t$), exceeds the scope of this Appendix.

\subsection{Pseudocode}
The computer codes used for generating numerical simulation results have been uploaded to online repositories \cite{code}, as stated on the main text. Here we summarize the code structure.
\\
\\
{Before network simulation starts, variables are defined.}
{
\color{blue}
\begin{verbatim}
**            Define variables              **
* Number of input, bulk and output nodes:    *
N_in, N_bulk,  N_out
* Algorithm Voltages:                        *
V_read=0.00001 V_write=-0.2
* Node voltage vector variables:             *
V_in(N_in), V_bulk(N_bulk), V_out(N_out)
* Resistance values and currents:            *
R_In-Bulk(N_in,N_out),
R_Bulk-Out(N_bulk,N_out),
I_In-Bulk(N_in,N_out),
I_Bulk-Out(N_bulk,N_out)
* Define other auxiliary variables           *
 \end{verbatim}}
 
{Each memristor parameter is chosen.}

{
\color{blue}
\begin{verbatim}
**    Generate memristor parameters         **
* beta, V_threshold, R_min/max               *
for i=1,N_in; for j=1,N_bulk
    beta_In-Bulk(i,j)=random(0.8-1)
    VT_In-Bulk(i,j)= random(0.05-0.1)
    R_min_In-Bulk(i,j)=random(50-100)
    R_max_In-Bulk(i,j)=5000
end for (j); end for (i)
* Do the same for beta_Bulk-Out,             *
* VT_Bulk-Out, and Rmin/max_Bulk-Out         *
 \end{verbatim}}
 
{A learning task is selected. In this case, a random input-output map is selected.}

{
\color{blue}
\begin{verbatim}
**             Generate random Map          **
for i=1,N_in
    input-output-MAP(i)=random_integer(1,N_out)
end for(i)
 \end{verbatim}}

{In the main routine, up to 1000 learning steps are performed. Within each learning step, up to $n_s=80 $ correction steps are applied.
The main routine uses three subroutines: READ, CORRECT and COMPUTE ERROR.}
\\
\\
\\
{
\color{blue}
\begin{verbatim}
***           Main Routine                 ***
for learning_step=1,1000
    input_node=random_integer(1,N_in)
    for correct<n_s
        call READ
           if (output_node =
                =input-output-MAP(input_node))
               FINISH current learning_step
           else 
              call WRITE(input_node,output_node)
    end for (correct)       
    COMPUTE ERROR
    if (error==0) SUCCESS, EXIT.
end for (learning_step)
END 
 \end{verbatim}}
 
 {
This is the pseudocode for READ, which requires the calculation of currents and voltages following Kirchoff's equations.
}
{
\color{blue}
\begin{verbatim}
READ routine(input_node)
for k=1,N_out
     Solve the circuit equations when 
     V_write is applied among input_node
     and the k-th output node.
end for (k)
Report output_node as the one 
with maximum current.
 \end{verbatim}}
 {
This is the pseudocode for WRITE.
}
{
\color{blue}
\begin{verbatim}
WRITE routine(input_node,output_node)
for time=1,5
    Solve the circuit equations when 
    V_read is applied among input_node 
    and output. 
    Calculate voltage difference on
    each memristor.
    Update resistances (using 
    memrisor equations).
end for(time)
 \end{verbatim}}
 {
This is the pseudocode for COMPUTE ERROR.
}
{
\color{blue}
\begin{verbatim}
COMPUTE ERROR routine
error=0
for i=1,N_in
    call READ(i)
    if (output_node /= input-output-MAP(i)) 
       error++
end for(i)
 \end{verbatim}}
 

\begin{thebibliography}{50}
\bibitem{units} In the following, it is customary to  express all resistances  in terms of an arbitrary unit resistance $R_0$, thus omitting units. Similarly, all voltages are written in terms of a unit voltage $V_0$, and time is also measured in terms of time unit $t_0$. Consequently, currents are measured in units of $V_0/R_0$, and parameter 
$\beta$ (which describes the rate at which memristor's resistance change, see the Appendix) in units of $R_0/V_0/t_0$.
\end{thebibliography}

\end{document}